# Lead-Free Perovskites

May Ming

North Carolina School of Science and Mathematics

1219 Broad Street, Durham, NC, 27707

# 1 Abstract

One of the most viable renewable energies is solar power because of its versatility, reliability, and abundance. In the market, a majority of the solar panels are made from silicon wafers. These solar panels have an efficiency of 26.4% (NREL, 2023) and can last more than 25 years (Sunrun, 2023). The perovskite solar cell is a relatively new type of solar technology that has a similar maximum efficiency and much cheaper costs, the only downside is that it is less stable and the most efficient type uses lead. The name perovskite refers to the crystal structure with an $ABX_3$ formula of the perovskite layer of the cell. All materials possess a property called a "band gap". The smaller the band gap the more conductive the material, but this does not necessarily mean that the smaller the band gap the better the solar cell. The Shockley-Queisser limit provides the optimal band gap in terms of efficiency for a single junction solar cell which is 1.34 eV for single junction cells (Shockley W., 1961). This research focuses on tuning the band gap of lead-free perovskites through B-site cation replacement. Through this investigation, the optical band gaps of tin and lead perovskites were re-established. However, the copper-based perovskite disagrees with existing DFT calculations. Additionally, the mixed tin and copper perovskite in this experiment contradicts the intuitive prediction.

# 2 Introduction

## 2.1 Why solar energy?

As the human population grows the demands for more energy will also increase. This will only cause an increase in greenhouse gasses, which will further increase the temperature of the Earth. Of course, to stop the increasing production of greenhouse gases there are two key components, one is reducing the amount of energy we consume, and the second is to use more renewable energies.

One of the most viable renewable energies is solar power. Solar power is versatile, cost-efficient, reliable, and abundant. In terms of versatility solar panels can be placed on almost any rooftop with access to sunlight. For cost-efficiency, solar panels need to provide the same amount or more of electricity in dollars than the initial cost of setting up solar panels. In terms of reliability, conventional solar panels made of silicon wafers can last more than 25 years (Wikipedia, 2023). For abundance, everywhere on Earth has access to the sun. Lastly, solar power could technically replace all energy if all the rooftops in the world (.2 million km$^2$) had solar panels installed (S Joshi, 2021). Finally, solar power is one of the most efficient and versatile of all renewable energies (Cleanmax, 2023).

**2.2 Why perovskite over other methods?**

In the market, a majority of the solar panels available are made from silicon wafers. These solar panels have an efficiency of 26.4% (NREL, 2023) and can last more than 25 years (Wikipedia, 2023). The perovskite solar cell is a relatively new type of solar technology that has a similar maximum efficiency, much cheaper costs, and the ability to be incorporated into more flexible technologies. These solar panels contain a layer of perovskite crystal, giving them their name. The downsides are that it is less stable, and the current leading perovskite in efficiency contains lead.

**2.3 Traditional Solar Cell Mechanics/p-n junction**

It is important to understand the underlying mechanics of traditional solar cells to understand this research's logic. A traditional individual solar cell, for example, Silicon wafer-based, can be represented, on a basic level, by a p-n junction. A p-n junction can be thought of as two semiconductors one p-type and one n-type. The p-type has holes or missing electrons while the n-type has excess electrons. The two sides begin with no net charge for each respective semiconductor. When you place the two types of semiconductors next to each other the electrons flow to the side with holes, and the holes migrate to the other side. Both charge carriers flow to the outer regions of their respective sides, leaving a depletion region in the middle. This creates a difference in charge between the two semiconductors. The difference in charge creates an electric field that opposes the direction of the flow of electrons and holes. If a p-n junction is just sitting the initial movement will occur, but eventually, equilibrium will be reached because of the opposing forces due to the electric field and the diffusion force. However, if a photon of light reaches the depletion region of the p-n junction it will provide enough energy for an additional hole and electron to split off. This furthers the difference in charge on both sides of the p-n junction creating a potential difference. When the p and n regions are connected through a wire the electrons are able to flow back to the n region, generating a current through the wire. There is also a voltage due to the difference in charges. If photons of light continue to hit the p-n junction this process will continue, which is how solar cells generate power.

**2.4 What is a n-i-p junction?**

A perovskite solar cell follows a n-i-p junction structure. A n-i-p junction is a pn junction with an intrinsic or undoped semiconductor layer between the p-type semiconductor and the n-type semiconductor. The intrinsic layer is similar to the depletion region in a p-n junction, it is where electron and hole pairs are created along with a potential difference, called the built-in potential difference (Superstrate, 2023). The p and n-type semiconductor still functions in the same way as a normal p-n junction.

**2.5 n-i-p Junction in the context of perovskites**

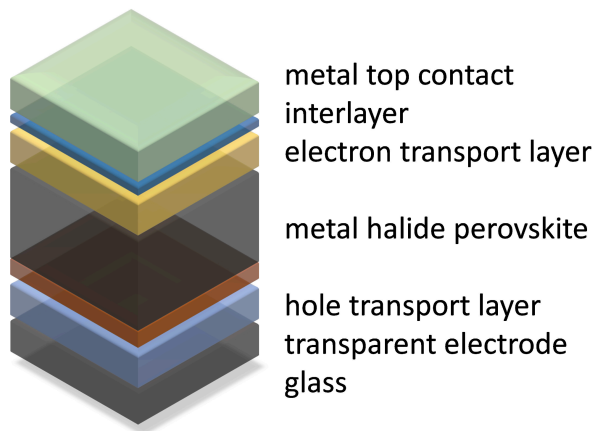

(M2N Research Group Eindhoven, 2022)

A perovskite solar cell is structured such that on the top there is a cathode (gold or silver), underneath is the hole transport layer (commonly Spiro-OMeTAD), underneath that is the intrinsic semiconductor (perovskite crystal), underneath that is the electron transport layer (often TiO$_2$), and the bottom layer is TCO or FTO coated glass. In perovskite solar cells the intrinsic layer is meant to absorb photons of light, which create electron and hole pairs. This project will only focus on the intrinsic semiconductor layer or perovskite crystal layer.

## 2.6 Semiconductors and band gaps

This research focuses on how changing the materials used as the semiconductor changes the bandgap which in turn changes how effective electricity generation is. Semiconductors are materials that are less conductive than conductors and more conductive than insulators. All materials possess a quality called a "band gap". A band gap is the difference in energy between the valence band and the conduction band in a material. The smaller the band gap the more conductive the material, but this does not necessarily mean that the smaller the band gap the better performing the solar cell. If the band gap is too large there can be thermalization loss, which is when the photon energy is lost as heat. If the band is too small there can be transmission loss, which is when the light passes through the material (Heidarzadeh, H., 2020). According to the Shockley-Queisser limit the ideal band gap for a single junction cell that reduces thermalization and transmission loss is 1.34 eV.

## 2.7 Issues with current perovskites

Currently, the leading perovskite in efficiency contains lead. Lead is found to be not only harmful to humans but also disruptive to the environment. Depending on the amount of lead in the human body, lead can cause damage to the nervous system, kidney, immune system, reproductive and developmental systems, or cardiovascular systems (EPA, 2023). Lead is most harmful to small children, leading to behavior problems, lower IQ, and learning deficits. As to the environment, Lead can easily leech into the soil and continue to leech into streams and lakes, only to be consumed by wildlife. The wildlife who consume high enough amounts of lead have a decrease in population growth rate and even neurological effects.

In the context of Solar Cells the European Union (RoHS2 from 2011) restricts the use of lead in electronics to be <0.1% of weight (EUR-Lex, 2011). Most Lead halide perovskites as of yet contain more than 10% of Lead by weight (M. Saliba, 2018). Even though the European Union made an exception for lead being used in solar cells it is still unideal to rely on lead-based technology. Additionally, lead is especially dangerous because it forms water-soluble lead compounds when brought into contact with water (Lenntech, 2023). Considering that solar cells are oftentimes found in an outdoor environment where it can be humid this creates more complications for lead-based perovskites.

## 2.8 Novel Lead-free perovskites

In recent years there has been research conducted on variations of perovskite solar cells that do not involve lead. Many of the substitute metals include Bi, Ge, Ag, and Sn (A. Abate, 2017). Tin is one of the most promising of the candidates; this past year, there was a paper published where an efficiency of 23.25% with Tin as the B-site cation was achieved (S. Ye et al.,2023). The method used in the more recent paper involved using a novel "full precursor" in place of the traditional half precursor (OGV - Energy, 2023).

## 2.9 Project Aims

Knowing that lead is not preferred for perovskites to be commercialized, the focus of this project will be exploring potential candidates for lead-free perovskite solar cells. According to previous research, $MASnI_3$ has a band gap of around 1.24 eV which is a little lower than the ideal band gap (Tao, S., 2019). According to another paper with a computational evaluation of $MACuI_3$, it has a band gap of around 1.25 eV (Gao, Z., 2022). In an experimental paper, the scientists created a mixed B-site cation perovskite with an equal ratio of copper and tin, achieving a band gap between just tin or just copper (Balachandran, N., 2021). In this project, multiple ratios will be tested to see if there is a trend and if the band gap can be tuned to be closer to ideal.

## 2.0.1 Perovskite Fabrication Techniques

Perovskites in contrast to conventional silicon wafer solar cells do not require vacuums or inert environments. Although less humid conditions do improve the performance of perovskites, they can still be easily fabricated at room temperature and in "normal" air. The design of a perovskite solar cell involves a substrate, coated in precursor material to form the perovskite crystal. In this project, because the goal is to measure the band gap of the thin film not the efficiency of a cell microscope glass will be used as the substrate.

# 3 Materials & Methods

## 3.1 Experimental Design

In this experiment, the metal used in the B-site cation of the perovskite was varied and the optical band gap as an effect of this was measured. The CH3NH3PbI3-based perovskite served as the control. The tin-based $CH_3NH_3SnI_3$ perovskite served as the first variation. For the second variation, a copper-based $CH_3NH_3CuI_3$ perovskite was used. Lastly, a mix of copper and tin perovskite was used for the last variation ($CH_3NH_3Sn_{1-x}Cu_xI_3$).

The materials used in this experiment include glassware to prepare the perovskite pre-cursor solutions, a glass stir rod to coat the substrate, a UV-Vis spectrometer, and a laptop to collect data.

**3.2 Methods**

Three compositions of perovskite thin films with single cation replacement were fabricated. The band gap of each was found experimentally through UV-visible spectroscopy. Afterwards, the mixed-cation thin films were fabricated and the band gap was found with the same procedure.

*3.2.1 Solutions*

First, the solutions were created by combining 140 mg $PbCl_{2\ (solid\ anhydrous)}$, 240 mg $CH_3NH_3I$, and 1 ml DMF, and stirring at 80C until a clear yellow solution was formed. The solution was then diluted in a 1 $MAPbI_3$:2 DMF ratio.

*3.2.2 Film Fabrication*

The film fabrication procedure followed that of Stoumpus et. al and Yokoyam et. al (C. C. Stoumpos, 2013), (T. Yokoyama et al., 2016).

For $MAPbI_3$, the glass substrate was placed on the hot plate with the precursor solution at 90°C. Once preheating was complete, one drop of perovskite precursor solution was placed on the clean piece of microscope glass and spread across the surface with a glass pipet. The pipet was rolled over the surface a few times to obtain a uniform smooth thin film. After the film turned black, the film was left on the heat for 10 minutes before the heat was turned off.

For MASnI3, the sample slide and solution were placed on the hot plate, and it was set to 80°C. After the preheating was completed, one drop of the perovskite precursor solution was placed on a clean piece of microscope glass and spread across the surface with a pipet. It was rolled over the surface several times to obtain a uniform, smooth, thin film. After the film turned black, the film was left on the hot plate for 30 minutes before turning the heat off.

For MACuI$_3$, the same procedure was followed as MASnI$_3$.

*3.2.3 Characterization*

UV-Vis spectrometry was used to characterize the films in this project. LoggerPro was used as a software to access the data from the Vernier Go Direct SpectroVis Plus spectrometer. A clean piece of microscope glass was used as the reference by placing the cut glass into the cuvette holder such that the glass is held perpendicular to the light. After calibrating the reference slide was replaced with the microscope slide with the thin film. The measurement was then recorded and this process was repeated for each sample.

*3.2.4 Creating Tauc Plot*

$$\alpha h\nu = A(h\nu - E_g)^n \qquad\qquad E = \frac{hc}{\lambda}$$

Equation 1                                              Equation 2

The Tauc plots were created following the method outlined in Makuła et. al (P. Makuła, 2018). The absorbance spectrum was used to derive the Tauc plot. Using Tauc's equation the absorbance coefficient ($\alpha$) can be found as a function of as a function of the absorbance. The energy of a certain wavelength can be found through Planck's equation. If it is a direct band gap then n is ½. If it is an indirect band gap then n is 2. Then plot ($\alpha$*energy)^2 or ½ against energy. Afterward, the tangent line to the point with the steepest slope was extrapolated and the x-intercept is the band gap.

**4 Results**

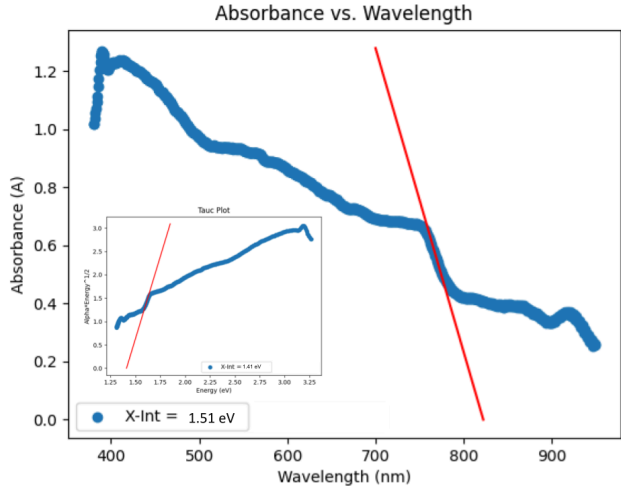 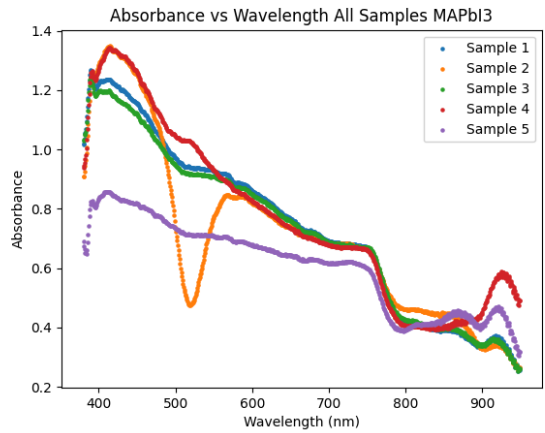

Figure 1: MAPbI$_3$ Sample 1 Absorbance vs. Wavelength with tangent line and a Tauc plot

Figure 2: Absorbance vs. Wavelength for all 5 MAPbI$_3$ samples

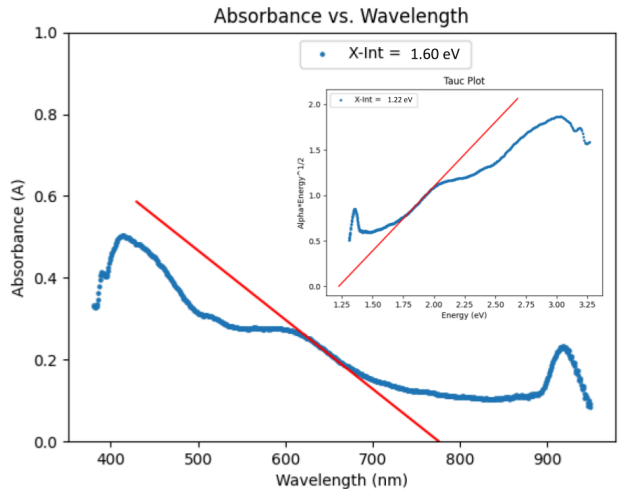 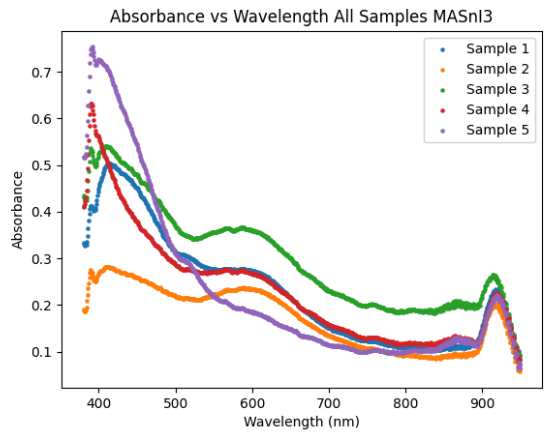

Figure 3: MASnI$_3$ Sample 1 Absorbance vs. Wavelength with tangent line and a Tauc plot

Figure 4: Absorbance vs. Wavelength for all 5 MASnI$_3$ samples

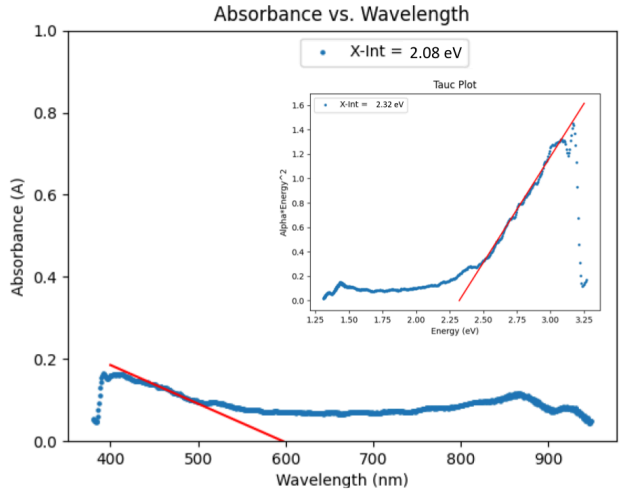

Figure 5: MACuI$_3$ Sample 1 Absorbance vs. Wavelength with tangent line and a Tauc plot

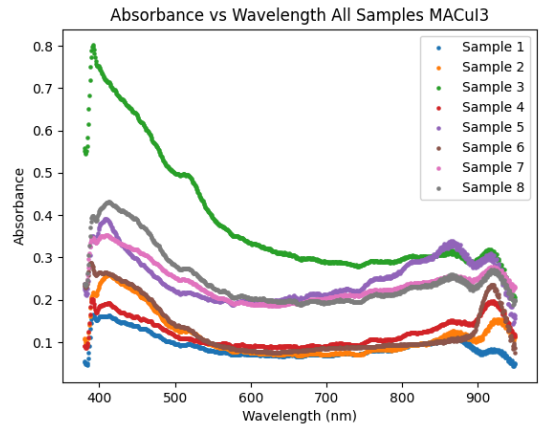

Figure 6: Absorbance vs. Wavelength for all 8 MACuI$_3$ samples

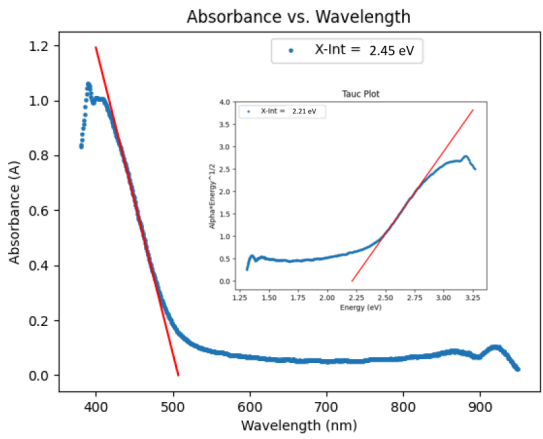

Figure 7: Absorbance vs wavelength plot (Cu:Sn) 1:2

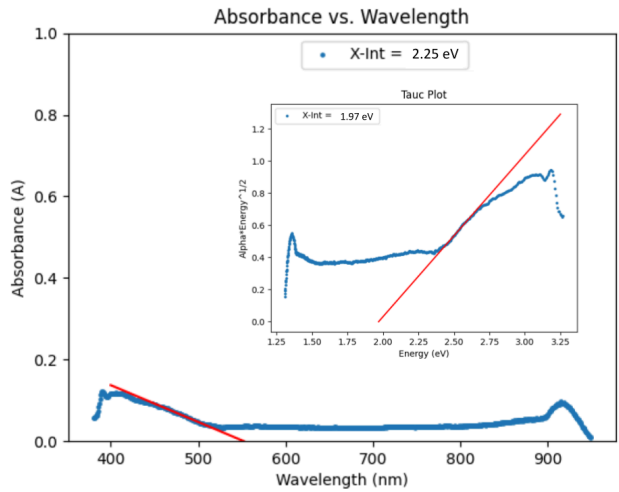

Figure 8: Absorbance vs wavelength plot (Cu:Sn) 1:1

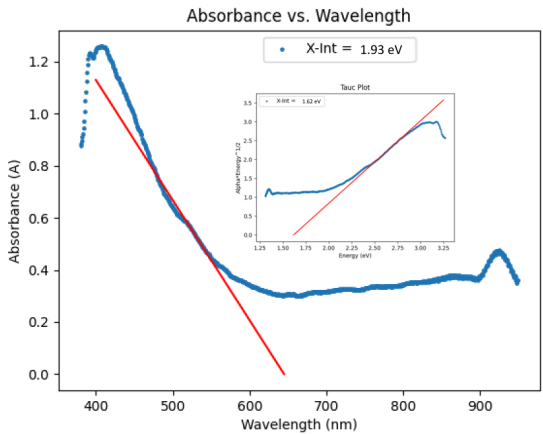

Figure 9: Absorbance vs wavelength plot (Cu:Sn) 2:1

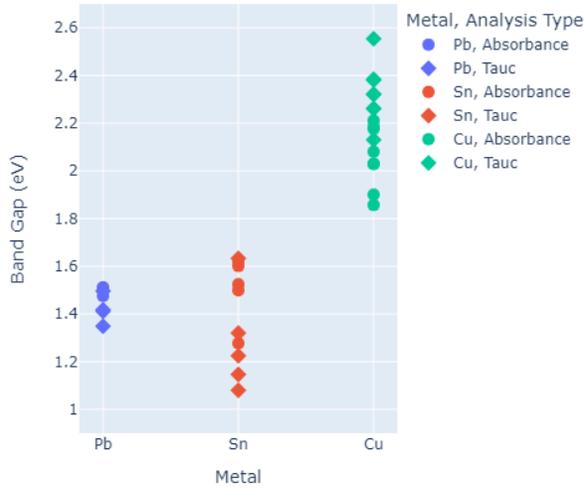
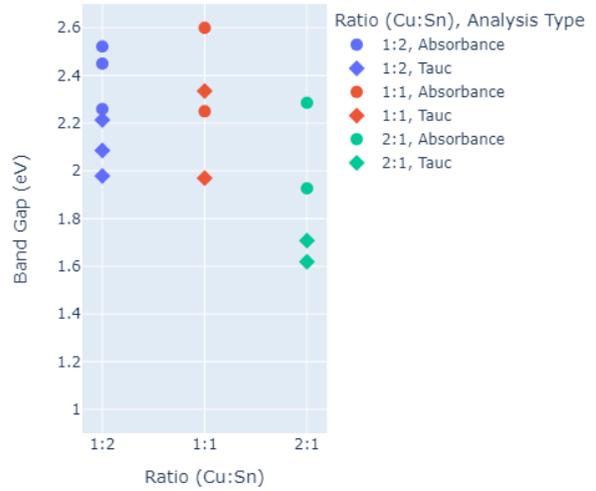

Figure 10: Dot plot of band gaps of the three different metals

Figure 11: Dot plot of band gaps of the three different ratios

| Metal | Average Band Gap (eV) | Average Band Gap from Previous DFT Calculations (Zhang C., 2020) (eV) |
|---|---|---|
| Pb | 1.46 | 1.59 |
| Sn | 1.39 | 1.52 |
| Cu | 2.19 | 1.25 |

Table 1: Average Band Gap of the 3 metals

| Ratio (Cu:Sn) | Average Band Gap (eV) |
|---|---|
| 1:2 | 2.31 |
| 1:1 | 2.27 |
| 2:1 | 1.88 |

Table 2: Average Band gap of 3 different ratios

# 5 Discussion

### 5.1 Single metal replacement

From Figure 10 it appears that on average MASnI$_3$ has the lowest band gap but with a significant amount of variability. MAPbI$_3$ has the next largest band gap with very little variability, and MACuI$_3$ has the largest band gap with the largest variation. The lead and tin optical band gaps found in this investigation follow the trend from previous DFT calculations as seen in Table 1, in addition to being close in absolute value. Copper however does not follow the trend; compared to another paper reporting 1.24 eV (Zhang, C., 2020) as a computationally found value, MACuI$_3$ in this investigation is significantly higher at 2.19 eV. Considering that there is oftentimes an underestimation of the band gap when using DFT (Perdew J., 1985) it is possible that the computational paper simply underestimated the band gap. However the difference is quite large, and in the same computational paper, the band gap for MAPbI$_3$ is 1.59 eV which is very close to the results of this project as seen in Table 1. A possible reason for this could be the dihydrate copper chloride in the anhydrous copper chloride.

**5.2 Mixed cation tuning**

From Figure 11, there appears to be a decreasing trend in bandgap as the concentration of the copper solution increases. This is not what you would expect because MASnI$_3$ has a lower band gap compared to MACuI$_3$ in this experiment. However from the computational paper mentioned earlier MACuI$_3$ had a predicted band gap of 1.25 eV while MASnI$_3$ had a band gap of 1.52 eV (Zhang C., 2020). This trend agrees with what one would expect if MACuI$_3$ had the same band gap as the DFT calculations, however, in this experiment MACuI$_3$ had a band gap of 2.19 eV, not 1.25 eV. Another note during the creation of the solution there were particles or crystals in the solution after the copper and tin solutions were mixed. Both started with no precipitates, and they only appeared after they were mixed. In the future, different approaches to mixing the metals should be tested as this method does not account for the difference in moles per unit volume. Additionally, the band gap of the three samples is based on a single measurement and not an average. In the future, it is important to fabricate multiple to eliminate the possibility of a coincidence.

**6 Conclusion**

In this investigation, MAPbI$_3$, MASnI$_3$, MACuI$_3$, and MASn$_{1-x}$Cu$_x$I$_3$ perovskite thin films were fabricated and the band gaps of the samples were found. The band gaps of MAPbI$_3$ and MASnI$_3$ followed an expected trend and were within .2 eV of the expected band gap. The band gap of MACuI$_3$ however, did not agree very well with another DFT calculation, instead, the band gap is nearly twice as large as expected. Although the computational calculations suggest that MACuI$_3$ would have a smaller band gap, the MACuI$_3$ perovskites investigated in this study suggested that it should be far larger. Going with that result the second part of this experiment involving mixed cation perovskites did not follow the predicted trend. Instead of the band gap increasing as the concentration of MACuI$_3$ increased the band gap decreased. Having three ratios might not be sufficient to establish this trend and will require further testing in the future. In addition to further testing of MASn$_{1-x}$Cu$_x$I$_3$, the author also intends to do further research on MACuI$_3$ fabrication to assess whether or not there is a reason the experimental band gap does not agree with the DFT calculations.

This research re-establishes the known experimental band gaps for MASnI$_3$ and MAPbI$_3$ and indicates a need for more research into MASn$_{1-x}$Cu$_x$I$_3$ for potential band gap tuning capabilities and other possible benefits of mixed B-site cation perovskites. This investigation focuses solely on the optical band gap of these novel perovskites. In the future, it is important to evaluate the stability and PCE of these perovskites to gain a broader understanding of these crystals.

## 7 Acknowledgements

I would like to thank Dr. Falvo, the Burroughs Wellcome Fund (BWF), and the NCSSM Foundation.